\documentstyle[12pt]{article}
\vspace{-2cm}
\begin{document}

\title{\large\bfseries
THE PHYSICAL MODEL OF SCHRODINGER ELECTRON\,.\\
SCHRODINGER CONVENIENT WAY FOR DESCRIPTION \\
OF ITS QUANTUM BEHAVIOUR. }
\vspace{-3cm}
\author{Josiph Mladenov Rangelov ,\\
Institute of Solid State Physics\,,\,Bulgarian Academy of Sciences \\
72\,Tsarigradsko chaussee\,\,1784\,\,Sofia\,,\,Bulgaria\,. }
\vspace{-1.2cm}
\date{}
\maketitle

\begin{abstract}
 The physical model (PhsMdl) of a Schrodinger nonrelativistic quantized
electron (SchEl) is built by means of a transition of the quadratic
differential particle equation of Hamilton-Jacoby (QdrDfrPrtEqtHam/Jkb)
into the quadratic differential wave equation of Schrodinger (QdrDfrWvEqtSch) in
this work, which interprets the physical reason of its quantum (wave and
stochastic) behaviour (QntWvBhv) by explanation of the physical reason
which forces the classical Lorentz electron (LrEl) to participate in
Furthian quantized stochastic oscillation motion (FrthStchOscMtn), which
turn it into quantum SchEl. It is performed that this transition is
realized by my consideration the Bohm's quantum potential as a kinetic
energy of the forced FurthStchOscMtn of the SchEl's well spread (WllSpr)
elementary electric charge (ElmElcChrg) close to a smooth thin trajectory
of a classical LrEl. There exist as an essential analogy between the
Furthian quantum stochastic trembling oscillation motion and the Brownian
classical stochastic trembling motion so and between the description of
their behaviours.

\end{abstract}

The object of this paper is to discuss and to bring a green light on the
problems of the physical interpretation of the nonrelativistic quantized
behaviour of the Schrodinger electron (SchEl), described by means of the
nonrelativistic quantum mechanics (NrlQntMch) laws and its mathematical
results. The purpose of the present work is to describe the felicitous
physical model (PhsMdl) of the SchEl. An obvious physical model (PhsMdl)
of the nonrelativistic quantized SchEl is built by means of some simple
mathematical transformation of the known classical quadratic differential
particle equation of Hamilton-Jacoby (ClsQdrDfrPrtEqtHam-Jkb) into the
quantum quadratic differential wave equation of Schrodinger (QntQdrDfrWv
EqtSch). After well physical substantiation it is performed that this
transformation is realized by taking in a consideration the so called
Bohm's quantum potential as a kinetic energy, what it is in reality,
of the forced Furthian stochastic oscillation motion (FurthStchOscMtn) of
the SchEl's well spread (WllSpr) elementary electric charge (ElmElcChrg)
within the nearness to the trajectory of the classical Lorentz' electron
(LrEl). In such a natural way this transition interprets the physical reason,
exciting the quantum (stochastic corpuscular-wave) behaviour (QntWvBhv) by
explanation of the physical reason which forces the classical LrEl to
participate in a quantized FrthStchOscMtn, which turns it into the quantized
SchEl. In this fashion the QntQdrDfrWvEqtSch is obtained through addition of
the kinetic energy of the SchEl's FrthStchOscMtn, expressed the dispersion of
its momentum or stochastic velocity $u$, to the ClsQdrDfrPrtEqtHam-Jkb.
Therefore the nonrelativistic quantized Furthian random trembling circular
oscillation motion with various radius values inside of different planes
could be roughly determined with some mathematical calculation by means of
the classical probabilities laws of both the classical stochastic theory
(ClsStchThr) and Maxwell classical electrodynamics (ClsElcDnm).

Hence the resonant electric interaction (ElcIntAct) of the SchEl's WllSpr
ElmElcChrg with the averaged electric intensity (ElcInt) of the StchVrtPhtns
from the fluctuating vacuum (FlcVcm)(zero-point radiation field) determines
the influence of its behavior because they creates their stochastically
diverse harmonic circular oscillations with various radii within the
neighborhood of the smooth narrow path of the LrEl, spreading and turning it
into some wide rough cylindrically spread path and inducting the transition
of the classical LrEl into the quantized SchEl. The ElcInt between the
WllSpr ElmElcChrg of the SchEl and the ElcInt of the resultant quantized
electromagnetic field (QntElcMgnFld) (zero-point field),determined by both
the boundary conditions and the existent StchVrtPhtns, forced the WllSpr
ElmElcChrg to participate within isotropic stochastically orientated in
the three-dimensional space circular oscillation, the averaged kinetic
energy, which every SchEl could obtained from the FlcVcm, may be obtained
by the following formula, well known from NrlClsMch :
\begin{equation}\label{ij}
\,E_k\,=\,\frac{m\,\langle(\omega)^2\,(\delta r)^2\,\rangle}{2}\,=
\,\frac{e^2}{\pi}\,\frac{m.C}{\hbar}\,\left\{\frac{\hbar}{m.C^2}\right\}^2
\,\int\limits_{\omega_{min}}^{\omega_{max}}\,\omega \,d\omega
\end{equation}

 As usual we suppose that the upper limit $\omega_{max}$ is equal to
double value of the energy at rest of the SchEl ($\omega_{max}\,=\,2 m.C^2$). As the
contribution of the lower limit $\omega_{min}$ has negligible importance, we
could suppose that $\omega_{min}\,=\,0$. In this approximation we cam easily
obtained from eq.(\ref{ij}) its following presentation :
\begin{equation}\label{ik}
\,E_k\,=\,\frac{2}{\pi}\,.\,\frac{e^2}{C.\hbar}\,.\,m.C^2
\end{equation}

 Hence the existence of the isotropic three-dimensional nonrelativistic
Furthian QntStchBhv of the SchEl within the nonrelativistic quantum mechanics
(NrlQntMchn) very strongly remind us about the classical StchBhv (ClsStchBhv)
of some Brownian stochastic particle (BrnStchPrt). Thence the ElcIntAct of
the SchEl's WllSpr ElmElcChrg (or a MgnIntAct of the neutron's MgnDplMm) with
the resonantly averaged ElcInt (or MgnInt for neutral massive hadron) of the
QntElcMgnFld of the existent StchVrtPhtns in the FlcVcm corresponds to the
stochastic action of the fluctuating resultant force on account of many
molecular impacts upon the BrnStchPrt at a time of its scattering. In our
PhsMdl of the SchEl we explain its FrthQntStchBhv and one assist sorting
the matter out the physical opinion of its parameter within the NrlQntMchn.

 In above elaborate we have possibility to present the spatial distribution
$\Upsilon\,(\varrho)\,$ of the ElcChrg of the WllSpr ElmElcChrg by dint of
Kirchoff's presentation of $\delta (\varrho)$-function :
\begin{equation}\label{il}
\,F\,(\varrho)\,=\,\left\{\frac{2}{3\,\pi}\right\}^{3/2}
\,\left\{\frac{m.C}{\hbar}\right\}^3\,\exp{\{-(\frac{\varrho}{\lambda_o})^2\}}
\end{equation}

 Here we must point that the spatial distribution (\ref{il}) of the SchEl's
WllSpr ElmElcChrg is caused by the participation of the Dirac's electron's
fine spread (FnSpr) ElmElcChrg in the isotropic three-dimensional
relativistic quantized Schrodinger's self-consistent strong correlated
fermion harmonic oscillation motion.The isotropic three-dimensional
relativistic quantized Schrodinger's self-consistent strong correlated
fermion harmonic oscillation motion of the FnSpr ElmElcChrg of the DrEl
may be correctly described by the three $\alpha $ ($\gamma$) matrixes of
four order. but in approximation of change of the strongly correlated
fermion harmonic oscillation by the incorrelated boson harmonic oscillation,
we could used the well-known orbital wave function (OrbWvFnc)
$\psi_o\,(\varrho) $ of the three-dimensional harmonic oscillator in its
ground state, having the following analytical presentation :
\begin{equation}\label{im}
\,\psi_o\,(\varrho)\,=\,\{\lambda_o \sqrt{\pi}\}^{-\frac{3}{2}}
\,\exp{\{-\frac{\varrho^2}{2\lambda_o^2}\}}
\end{equation}

where $\lambda_o$ is the constant of the oscillation ($\lambda_o^2\,=
\,\frac{\hbar}{m\,\omega}\,=\,\frac{3}{2}\,.\{\frac{\hbar}{m.C}\}^2\,=
\,\frac{2}{3}\,.\,\langle\,\varrho^2\,\rangle\,$.  After some cursory
comparison it is easily to understand, that Kirchoff $\delta$-function
$F\,(\varrho)$ (\ref{il}) is obtained from the OrbWvFnc $\psi_o\,(\varrho)$
(\ref{il}) by means of the equation $F\,(\varrho)\,=\,\left|\,\psi_o\,
(\varrho)\,\right|^2$, the well known from the NrlQntMch.

 In order to obtain the averaged potential of the SchEl we must put into
right side of Poison equation the spatial distribution of ElcChrg
$\Upsilon\,(\varrho)\,$ of its WllSpr ElmElcChrg.In such a naturally way
we have possibility to calculate roughly the averaged self-potential of
the SchEl's WllSpr ElmElcChrg and to obtain its following excertional
presentation :
\begin{equation}\label{in}
\,V\,(\varrho)\,=\,\frac{-2.e}{\sqrt{\pi}.\varrho}
\,\int_o^{(\frac{\varrho}{\lambda_o})}\,\exp{(-x^2)}\,dx
\end{equation}

 The potential energy of the electric self-action (ElcSlfAct) of the
SchEl's WllSpr ElmElcChrg with the spatial distribution of its ElcChrg
$\Upsilon\,(\rho)\,$ from (\ref{il}) and its own potential $V\,(\rho)$ from
(\ref{in}) we are capable to determine by means of the following obvious
presentation :
\begin{equation}\label{io}
\,E_p\,=\,\frac{2e}{\sqrt{\pi}}\,.\,\frac{4\,\pi\,e}{\pi\sqrt{\pi}\lambda_o}\,
\int_o^\infty\,\exp(-u^2)\,\frac{u^2\,du}{\,u\,}\,\int_o^u\,\exp(-x^2)\,dx
\end{equation}

 The twofold integration can be easily execute by integration by parts.
In such a way after elementary calculation we can obtain the following
 obvious result :
\begin{equation}\label{ip}
\,E_p\,=\,\frac{4\,e^2}{\lambda_o\pi}\,\int_o^\infty\,\exp(-x^2)\,dx\,=
\,\sqrt{\frac{2}{\pi}}\,\frac{e^2}{\lambda_o}\,=
\,\frac{2}{\sqrt{3\pi}}\,.\,\frac{e^2}{C.\hbar}\,.\,m.C^2
\end{equation}

 In further after some cursory comparison of the eqs.(\ref{ik}) and
(\ref{ip}) we could understand that the value of the averaged kinetic
energy, which the SchEl obtain from the FlcVcm at its stochastic circular
oscillations, is equal of the value of the potential energy of its
ElcSlfAct between spatial density of its WllSpr ElmElcChrg and its own
averaged potential. This equality is no accidental nature and for certain
have important significant. After this comparison we can understand why
the potential energy of the own averaged electric potential has no
contribution into the rest energy of the DfEl. It turns out that every
electron obtains the potential energy of the ElcSlfAct of its WllSpr
ElmElcChrg by its own averaged potential in form of the kinetic energy
on account of its participation in the isotropic three-dimensional
nonrelativistic quantized stochastic boson harmonic oscillations from the
FlcVcm at the interaction of its WllSpr ElmElcChrg with the ElcInt of the
StchVrtPhtns. Moreover, we can easily understand that the participation
of the WllSpr ElmElcChrg of the SchEl in the isotropic three-dimensional
nonrelativistic quantized stochastic boson harmonic oscillations not only
takes its illocalizing energy from the FlcVcm, ensuring with this the
stability of its ground state in H-atom, but at this as well as all this
oscillation create its additional MchMmn and MgnDplMmn, and this ElcIntAct
its tunnelling through some potential barriers and causes the shift of its
energy levels in atoms.Therefore all this experimental observed phenomena
in the long run demonstrate the real participate of the SchEl in the
isotropic three-dimensional nonrelativistic quantized stochastic boson
harmonic oscillations as a result of the ElcIntAct of its WllSpr ElmElcChrg
with the EctInt of the resultant QntElcMgnFld of the existent StchVrtPhtns.

Although till now nobody know what the McrPrt means, all the same there
exists a possibility for a consideration of an unusual behaviour of a
QntMcrPrt by means of a transparent surveyed PhsMdl of the SchEl. In our
PhsMdl the SchEl will be treated as a well spread (WllSpr) ElmElcChrg,
taking simultaneously part in two different motions: A/The classical motion
of the LrEl along an well contoured smooth and thin trajectory realized in
a consequence of some classical interaction (ClsIntAct) of its over spread
(OvrSpr) ElmElcChrg, bare mass or magnetic dipole moment (MgnDplMm) with
some external classical fields (ClsFlds), described by well known laws of
the Newton nonrelativistic classical mechanics (NrlClsMch). This motion may
be finically described by virtue of the laws of both the NrlClsMch and the
classical electrodynamics (ClsElcDnm); B/The isotropic three-dimensional
nonrelativistic quantized (IstThrDmnNrlQnt) Furthian stochastic boson
harmonic oscillation motion (FrthStchBznHrmOscMtn) of the SchEl as a result
of the permanent ElcIntAct of the electric intensity (ElcInt) of the
self-consistent resultant QntElcMgnFld of all the StchVrtPhtns, existing
within the FlcVcm and generated by dint of the VrtPhtn's stochastic exchange
between them. The SchEl's motion and its unusual quantized behaviour,
described in the NrlQntMch may be easily understood by assuming it as a
forced random trembling oscillation motion (RndTrmMtn) upon a stochastic
joggle influence of the StchVrtPhtns scattering from some BrnClsPrt.
Therefore the RndTrmMtn can be approximately described through some
determining calculations by means of both the laws of the Maxwell ClsElcDnm
and the probable laws of the classical stochastic theory (ClsStchThr).
But in a principle the exact description of the SchEl's uncommon behaviour
can be carry into a practice by means only of the NrlQntMch's laws and
ClsElcDnm s ones.

 In an accordance of the analogy between the Furthian quantum stochastic
trembling oscillation motion and the Brownian classical stochastic trembling
motion and the description of their behaviours (of the BrnClsPrts and of the
FrthQntPrts) with a deep physical understanding of the Furthian random
trembling oscillation motion (FrthRndTrmOscMtn), we must determine both as
the value $V_j^-$ of the BrnClsMcrPrt's (FrthQntMcrPrt) velocity before the
moment $t$ of the scattering time of some molecule (LwEnr-StchVrtPhtn) from
one (its OvrSpr ElmElcChrg), so the value $V_j^+$ of its velocity after the
same moment $t$ of the scattering time by means of the following
definitions\,:
\begin{equation}\label{a} \, V_j^-(r,t)\,=\,L i m_{Dt \to o}
\left(\frac{r_j (t) - r_j (t - Dt)} {Dt}\right)\,; \, V_j^+(r,t)\,=\,L i
m_{Dt \to o} \left(\frac{r_j (t.+ Dt) - r_j (t)} {Dt}\right)\,;
\end{equation}

 In addition we may determine two new velocities $v_j$ and $u_j$ by dint
of the following simple equations :
\begin{equation}\label{b}
\quad 2\,V_j\,=\,\left[\,V_j^+\,+\,V_j^-\,\right]\,;\quad,
\quad 2\,i\,U_j\,=\,\left[V_j^+\,-\,V_j^-\,\right]\,;\\
\end{equation}

 In conformity with the eqs.(\ref{b}) it is obviously followed that the
current velocity, having a real value $V$, in reality describes the regular
drift of the BrnClsMcrPrt (FrthQntMcrPrt) and the osmotic velocity, having a
imagine value $iU$, in reality describes nonrelativistic Brawnian classical
(Furthian quantized) stochastic trembling harmonic oscillations. Afterwards
by virtue of the well-known definition equations\,:
\begin{equation}\label{c}
 2m\,V_j\,=\,m\,\left[\,V_j^+ \,+\,V_j^-\,\right]\,=
\,2\,\nabla_j\,S_1\, ; \quad and \quad
\, 2i\,m\,U_j\,=\,m\,\left[\,V_j^+\,-\,V_j^-\,\right]\,=
\,2\,i\,\nabla_j\,S_2 \\, ;\\
\end{equation}

one can obtain following presentation of the SchEl's OrbWvFnc $\Psi(r,t)\,$ :
\begin{equation}\label{d}
\,\Psi(r,t)\,=\,\exp\left(\,\frac{ i S_1}{\hbar}\,-
\,\frac{ S_2}{\hbar}\,\right)\,=
\,B\,\exp\,\left(\,\frac{i S_1}{\hbar}\,\right)
\end{equation}

 It is easily to verify the results (\ref{c}) and (\ref{d}). In  effect one
be obtained by means of the following natural equations\,:
\begin{equation}\label{ea}
\,m\,V_j^+\,\Psi\,=\,-i\,\hbar\nabla_j\,\exp\,\left(\,\frac{-i S_1}{\hbar}\,-
\,\frac{S_2}{\hbar}\,\right)\,=
\,\left(\,\nabla_j\,S_1\,+\,i\,\nabla_j\,S_2\,\right)\,\Psi\quad ;\\
\end{equation}

\begin{equation}\label{eb}
\,m\,V_j^-\,\Psi^+\,=\,+i\,\hbar\nabla_j \exp\,\left(\,\frac{iS_1}{\hbar}\,-
\,\frac{S_2}{\hbar}\,\right)\,=
\,\left(\,\nabla_j\,S_1\,-\,i\,\nabla_j\,S_2\,\right)\,\Psi^+\quad ;\\
\end{equation}

 In this fashion the QntQdrDfrEqtSch is obtained through addition of the
kinetic energy of the SchEl's FrthRndTrmMtn, expressed with the dispersion
of its momentum or stochastic velocity, to the ClsQdrDfrEqtHam-Jkb. Hence
the classical motion of the LrEl is described by a smooth narrow path,
which is determined from its classical real part $S_1$ of the complex
action $S[r,t]$ and its derivatives, but the Furthian quantized stochastic
motion of the SchEl is described by a rough cylindrically spread broad path,
which is determined correctly from its imaginary part $S_2$ represented by
the module of its orbital wave function (OrbWvFnc) $\Psi(r,t)$ and operators.
Consequently, the quantum motion is described by rough broad path, which is
determined from the quantum action $S[r,t]$ and its derivatives by the
orbital wave function (OrbWvFnc) $\Psi(r,t)$ and operators. It turns out,
that if the action function $S(r,t)$ has only a real value $S_1$, then the
micro particle (McrPrt) moves along a classical well contoured smooth and
narrow path ; but when if the action function $S(r,t)$ has only imaginary
value $S_2$, then the McrPrt moves on a its trajectory, cylindrically
spread and turned into wide path of the cylindrical form with differ radii
and centers, being on small pieces from stochastically broken line ; when
the action function has a complex value $S(r,t)$, then the McrPrt moves in
the quantized dual form : indeed as the real part $S_1$ of the action
function and its derivative determine the classical motion and its current
velocity $v$ and the imaginary part $S_2$ of the complex action function
$S(r,t)$ and its derivative determine the stochastic motion and its osmotic
velocity $u$. This spread of the smooth thin curve through its cylindrically
spread and turned into wide path of the cylindrical form with differ radii
and centers, being on petty breaking of small pieces makes the trajectory
in rough and road path, which forces us to put the OrbWvFnc $\Psi(r,t)$
description of the SchEl's behaviour.Hence the classical motion of the LrEl
is described by a smooth narrow path, which is determined from its classical
real part $S_1$ of the complex action $S(r,t)$ and its derivatives, but the
Furthian stochastic quantum oscillation motion of the SchEl is described by
a wide rough path, which is mathematical correctly determined from its
imaginary part $S_2$ represented by the module of its orbital wave function
(OrbWvFnc) $\Psi(r,t)$ and operators. It turns out, when the action function
$S(r,t)$ has only a real value $S_1$, then the NtnMcrPrt moves along its
classical well contoured smooth and narrow path\,;\,when the action function
$S(r,t)$ has only an imaginary value $S_2 $, then the BrnMcrPrt moves
stochastically on a frequently broken and very scattered orientated line of
small pieces\,;\, when the action function $S$ has a complex value,\,then the
QntMcrPrt moves in a quantized dual form\,:\,as the real part $S_1$ of the
action function $S$ and its derivatives determine the classical motion and
its current velocity $v$ and the imaginary part $S_2$ of the action function
$S$ and its derivatives determine the forced stochastic motion and its
spreading (osmotic) velocity $u$. This spreading of the thin and smooth
classical trajectory through its wide path of the cylindrical form with
differ radii and centers, being on often breaking of small pieces forces us
to put the OrbWvFnc $\Psi(r,t)$ for description of the SchEl's behaviour.

 Indeed, it is well known that the imaginary part of the energy of the McrPrt
describes its decay in the time and the imaginary part of the velocity of the
McrPrt describes its going out from the classical trajectory in the space,
which is forbidden for the free motion of the ClsMcrPrt. Therefore the module
quadrate of the SchEl's OrbWvFnc $\Psi(r,t)$ $\mid\Psi(r,t)\mid^2$, where
hasn't any imaginary part (i.s.has no real part $S_1$ of its action function
$S$), describes only its probability for its discovering (location) in a very
small area of the space,close by the space point having coordinates r,in the
moment t of the time.The fluctuating alternation of the imaginary parts of
the SchEl's energy and momentum (quantities of motion) may be considered as
a result of continuous exchange of some parts of its energy and momentum at
the uninterrupted alternative absorption and emission of the stochastic
virtual photons (StchVrtPhtns) within the fluctuating vacuum (FlcVcm).

 In a consequence of what was asserted above in order to obtain the QntQdrDfr
WvEqn of Sch we must add to the kinetic energy $\,\frac{(\nabla_l\,S_1)^2}
{2m}\,$ of the NtnClsPrt in the following ClsQdrDifPrtEqt of Hml-Jcb
\begin{equation}\label{f}
 -\frac{\partial S_1}{\partial t}\,=\,\frac{(\nabla_j\,S_1)^2}{2m}\,+\,U\,;
\end{equation}

the kinetic energy $\,\frac{(\nabla_l\,S_2)^2}{2m}\,$ of the BrnClsPrt. In
such the natural way we obtain the following analytic presentation of the
QntQdrDfrWvEqt of Sch \,:
\begin{equation}\label{g}
 -\frac{\partial S_1}{\partial t}\,=\,\frac{(\nabla_j\,S_1)^2}{2m}\,+
\,\frac{(\nabla_j\,S_2)^2}{2m}\,+\,U\,;
\end{equation}

It is obviously to understand that the first term $\,\frac{(\nabla_l\,S_1)^2}
{2m}\,$ in the eq.(\ref{g}) describes the kinetic energy of the regular
translation motion of the NtnClsPrt with its current velocity $v_l\,=\,
\frac{\nabla_l\,S_1}{m}\,$ and the second term $\,\frac{(\nabla_l\,S_2)^2}
{2m}\,$ describes the kinetic energy of the random trembling oscillation
motion (RndTrmOscMtn) of the BrnClsPrt with its osmotic velocity $u_l\,=\,
\frac{\nabla_l\,S_2}{m}\,$ . Therefore we can rewrite the expression
(\ref{g}) in the following form\,:
\begin{equation}\label{h}
 \quad E_t\,=\,\frac{m\,v^2}{2}\,+\,\frac{m\,u^2}{2}\,+\,U\,= \,
\frac{\langle\,P\,\rangle^2}{2\,m}\,+\,\frac{\langle(\Delta\,P)^2\rangle}
{2\,m}\,+\,U\,;
\end{equation}

 After elementary physical obviously suppositions some new facts have been
brought to light. Therefore the upper investigation entitles us to make the
explicit assertion that the most important difference between the QntQdrDfr
WvEqt of Sch and the ClsQdrDfrPrtEqt of Hml-Jcb is exhibited by the existence
of the kinetic energy of the FrthRndTrmOscMtn in the first one. Therefore
when the SchEl is appointed in the Coulomb's potential of the atomic nucleus
fine spread (FnSpr) electric charge (ElcChrg) $Ze$ its total energy may be
written in the following form :
\begin{equation}\label{i}
\langle\,E_t\,\rangle\,=\,\frac{1}{2\,m}\,\left[\langle P_r \rangle^2\,+\,
\frac{\langle L \rangle^2}{\langle r \rangle^2}\,\right]\,+\,
\frac{1}{2\,m}\,\left[\langle(\Delta P_r)^2 \rangle\,+\,
\frac{\langle(\Delta L)^2 \rangle}{\langle r \rangle^2}\,\right]\,-\,
\frac{Z e^2}{\langle r \rangle}
\end{equation}

As any SchEl has eigenvalues $n_r\,=\,0\,$ and $l\,=\,0\,$ in a case of its
ground state, so it follows that $\langle P_r \rangle \,=\,0\,$ and
$\langle L \rangle\,=\,0\,$. As a consistency with the eq.(\ref{k}) the
eigenvalue of the SchEl's total energy $E_t^o$  in its ground state in some
H-like atom is contained only by two parts :
\begin{equation}\label{j}
\langle\,E_t^o\,\rangle\,=\,
\frac{1}{2\,m}\,\left[\langle(\Delta P_r)^2 \rangle\,+\,
\frac{\langle(\Delta L)^2 \rangle}{(\langle r \rangle)^2}\,\right]\,-
\,\frac{Z e^2}{\langle r \rangle}
\end{equation}

 Further the values of the dispersions $\langle (\Delta P_r)^2\rangle$ and
$\langle(\Delta L)^2\rangle$ can be determined by virtue of the Heisenberg
Uncertainty Relations (HsnUncRlt)\,:
\begin{equation}\label{k}
\,\langle (\Delta P_r)^2\rangle\,\times\,\langle (\Delta r)^2\rangle\,\ge
\,\frac{\hbar^2}{4}
\end{equation}
\begin{equation}\label{l}
\langle (\Delta L_x)^2\rangle\,\times\,\langle (\Delta L_y)^2\rangle\,\ge
\,\frac{\hbar^2}{4}\,\langle (\Delta L_z)^2\rangle\,
\end{equation}

 Thence the dispersion $\langle(\Delta P_r)^2\rangle$ will really have its
minimal value at the maximal value of the $\langle(\Delta r)^2\rangle\,=
=\,\langle r \rangle^2$.In this way the minimal dispersion value of the $
\langle(\Delta P_r)^2\rangle$ can be determined by the following equation :
\begin{equation}\label{m}
\,\langle(\Delta P_r)^2\rangle\,=\,\frac{\hbar^2}{4\langle r^2\rangle}\,
\end{equation}

 As the SchEl's ground state has a spherical symmetry at $l\,=\,0\,$, then
the following equalities take place :
\begin{equation}\label{n}
\,\langle(\Delta L_x)^2\rangle\,=\,\langle(\Delta L_y)^2\rangle\,=
\,\langle(\Delta L_z)^2\rangle\,;
\end{equation}

 Hence we can obtain minimal values of the dispersions (\ref{n}) through
division of the eq.(\ref{k}) with the corresponding equation from the eq.
(\ref{n}). In that a way we obtain the following result :
\begin{equation}\label{o}
\,\langle(\Delta L_x)^2\rangle\,+\,\langle(\Delta L_y)^2\rangle\,+
\,\langle(\Delta L_z)^2\rangle\,=\,\frac{3\hbar^2}{4}\;
\end{equation}

 Just now we are in a position to rewrite the expression (\ref{k}) in the
handy form as it is well-known :
\begin{equation}\label{p}
\,E_t^o\,=\,\frac{1}{2\,m}\,\left[\,\frac{\hbar^2}{4r^2}\,+
\,\frac{3\hbar^2}{4r^2}\,\right]\,-\,\frac{Z\,e^2}{r}\,=
\,\frac{1}{2}\,\frac{\hbar^2}{m\,r^2}\,-\,\frac{Z\,e^2}{r}\,;
\end{equation}

 Subsequently the minimal value of the $E_t^o$ may be determined by
minimization of the expression (\ref{o}) in respect of the radius r. In such
a way we could obtain the minimizing equality :
\begin{equation}\label{q}
\,\frac{\partial E_t^o}{\partial r}\,|_{r\,=\,r_o}\,=
\,\frac{-\hbar^2}{m\,r^3}\,+\,\frac{Z\,e^2}{r^2}\,=\,0\,;
\end{equation}

 Thence we can obtain the value of the SchEl's orbital radius r in its
ground state of an H-like atom as a result of the minimizing eq.(\ref{o})
\begin{equation}\label{r}
\quad r_o\,=\,\frac{\hbar^2}{2me^2}\,=\,\frac{a_o}{Z}\,;
\end{equation}

 Here $a_o$  is the Bohr's radius of the SchEl's ground state in the H-like
atom.Further we can obtain the averaged value of the SchEl's total energy
$E_t^o$ when it occupies its ground state by the substitution of the
following equilibrium value of the orbital radius r from the eq.(\ref{n})\,:
\begin{equation}\label{s}
\,\langle\,E_t^o\,\rangle\,=\,-\,\frac{m Z^2 e^4}{2\hbar^2}\,;
\end{equation}

 Since then it is easily to understand by means of upper account that if
he ClsMcrPrt s motion is going along the clear definitized smooth thin
trajectory in accordance with the NrlClsMch,then the QntMcrPrt's motion is
perform in the form of the RndTrbOscMtn rough broad roadway near classical one
of any NtnClsPrt within NrlClsMch. As a result of that we can suppose that
the unusual dualistic behaviour of QntMcrPrt can be described by dint of the
following physical quantities within NrlQntMch :
\begin{equation}\label{t}
\,r_j\,=\,\bar {r}_j\,+\,\delta {r}_j\quad;
\quad p_j\,=\,\bar {p}_j\,+\,\delta {p}_j\quad;
\end{equation}

 Indeed, because of existence of $\,\delta {r}_j \,\ne\,0\,$ and
$\,\delta {p}_j \,\ne\,0\,$ within the NrlQntMch the value of the MchMm's
square $\,\langle L^2\rangle\,$ of the SchEl is different from the value of
averaged MchMmn's square $\,\langle L \rangle^2\,$ of the LrEl in the
NrlClsMch. Really,by dint of the Heisenberg Commutation Relations
(HsnCmtRlt)\,:
\begin{equation}\label{u}
\,L_x \,L_y \,- \,L_y \,L_x \,=\,i\,\hbar\,L_z\,;
\,L_y \,L_z \,- \,L_z \,L_y \,=\,i\,\hbar\,L_x\,;
\,L_z \,L_x \,- \,L_x \,L_z \,=\,i\,\hbar\,L_y\,;
\end{equation}
we can write two analogous inequalities : the inequality (\ref{l}) and
the following corresponding inequality :
\begin{equation}\label{v}
\langle(\Delta L_y)^2\rangle\,\times\,\langle(\Delta L_z)^2\rangle\,\ge
\langle(\Delta L_x)^2\rangle\quad;\quad
\langle(\Delta L_z)^2\rangle\,\times\,\langle(\Delta L_x)^2\rangle\,\ge
\langle(\Delta L_y)^2\rangle\,
\end{equation}

 We can suppose in following that in a case when the SchEl is placed in the
external potential of cylindrical symmetry its MchMn's component along the
axis Z has averaged value$\,<L_z>\,=\,l\hbar$ In a spite of that the averaged
value of the MchMn's square must be determined by the following equality :
\begin{equation}\label{w}
\,\langle L^2 \rangle\,=\,(\langle L_z\rangle)^2\,+
\,\langle(\Delta L_x)^2\rangle\,+
\,\langle(\Delta L_y)^2\rangle\,+\,\langle(\Delta L_z)^2\rangle\,;
\end{equation}

 Further the values of the quantities $\langle(\Delta L_x)^2\rangle\,$
$\langle(\Delta L_y)^2\rangle\,$ and $\langle(\Delta L_z)^2\rangle\,$
can be determined by virtue of the inequalities (\ref{u}) and (\ref{v})
in the following form :
\begin{equation}\label{z}
\,\langle(\Delta L_x)^2\rangle\,=\,\langle(\Delta L_y)^2\rangle\,=\,
\frac{l\hbar^2}{2}\quad and
\quad\langle(\Delta L_z)^2\rangle\,=\,\frac{\hbar^2}{4}
\end{equation}

 Then it is quite naturally that we must obtain the averaged value of the
MchMn's square at experiment,which is well-founded by my physical point of
view:
\begin{equation}\label{ab}
\,\langle L^2\rangle\,=\,l^2\hbar^2\,+\,\frac{l\hbar^2}{2}\,+
\,\frac{l\hbar^2}{2}\,+\,\frac{\hbar^2}{4}\,=
\,\hbar^2\,(l\,+\,\frac{1}{2})^2\
\end{equation}

 I think my successful picturesque example illustrates very exactly the
extraordinary situation of the QntMcrPrt within the NrlQntMch. Hence the
difference between the NtnClsBhv of the NtnClsMcrPrt, described by the laws
of the NtnClsMch, the BrnStchBch of the BrnClsMcrPrt, described by the laws
of the ClsStchMch, and the FrthStchBhv of the FrthQntMcrPrt, described by the
laws of the NrlQntMch may be roughly understand by means of three different
values of the action function $S$. It turns out, when the action function
$S(r,t)$ has only a real value $S_1$, then the NtnMcrPrt moves along its
classical well contured smooth and narrow path\,;\,when the action function
$S(r,t)$ has only an imaginary value $S_2 $, then the BrnMcrPrt moves
stochastically on a frequently broken and very scattered orientated line of
small pieces\,;\, when the action function $S$ has a complex value,\,then the
QntMcrPrt moves in the quantized dual form\,:\,as the real part $S_1$ of the
action function $S$ and its derivatives determine the classical motion and
its current velocity $v$ and the imaginary part $S_2$ of the action function
$S$ and its derivatives determine the forced stochastic motion and its
spreading (osmotic) velocity $u$. This spreading of the thin and smooth
classical trajectory through wide path of the cylindrical form with differ
radii and centers, being on often breaking of small pieces forces us to put
the OrbWvFnc $\Psi(r,t)$ for description of the SchEl's behaviour.

\newpage

     R E F E R E N C E S  \\
\vspace{1cm}

1.Rangelov J.M.,Reports of JINR ,$R4-80-493\,;\,R4-80-494,\,(1980)$,Dubna \\
2.Rangelov J.M.,University Annual (Technical Physics),$\underline{22} ,
\,(2),\,65,\,87,\,(1985)\,; \\
\underline {23} ,\,(2),\,43,\,61,\,(1986)\,;\underline {24} ,\,(2),
\,287,\,(1986)\,.$ \\
3.Rangelgov J.M.,Comptes Rendus de l'Academie Bulgaries Sciences,
$\underline{39} ,\,(12),\,37,\,(1986)$ \\
4.Rangelov J.M.,University Annual (Technical Physics),$\underline{25} ,
\,(2),\,89,\,113,\,(1988).$\\
5.De Broglie L.,Comptes Rendus $\underline{177} ,507,\,548,\,630,\,(1923).$\\
6.Heisenberg W.,Mathm.Annalen $\underline{95} ,694,(1926);$ Ztschr.f.
Phys.$\,\underline{33}\,,\,879,\,(1925)\,;\,\underline{38} ,\,411,\,(1926).$\\
7.Pauli W.,Ztschr.f.Phys.,$\underline{31} ,\,765,\,(1925)\,; \underline{36} ,
\,336,\,(1926)\,;\underline{41} ,\,81,\,(1927)\,.$\\
8.Schrodinger E.,Annal d. Phys.$\underline{79} ,\,361,\,489\,;\underline{80} ,
437\,;\underline{81} ,\,109,\,(1926)\,.$\\
9.Born M.,Heisenberg W.,Jordan P.,Ztschr.f.Phys.,
$\underline{35} ,\,557,\,(1926)\,.$\\
10.Dirac P.A.M.,Proc.Cambr.Phil.Soc.$\underline{22} ,\,132,\,(1924)\,;$ \\
Proc.Roy.Soc.A,$\underline{106} ,\,581,\,(1924)\,; \underline{112}
,\,661,\,(1926).$\\
11.Madelung E.,Ztschr.f.Phys.$\underline{40} ,\,322,\,(1926)\,$\\
12.Furth R.,Ztschr.f.Phys.$\underline{81} ,\,143,\,(1933)\,.$\\
13.Dirac P.A.M., Proc.Roy.Soc.A,$\underline{117} ,\,610\,; \underline{118}
,\,351,\,(1928)\,.$\\
14.Breit D.,Proc.Nat.Acad.Scien.USA,$\,\underline {14} ,\,553,\,(1928)\,;
\underline{17} ,\,70,\,(1931)\,.$ \\
15.Fock V.A., Ztschf.f. Physik,$\underline{55} ,127.(1928);
\underline{68} ,\,527,\,(1931)\,.$ \\
16.Wiener N.,Jour.Mathm.Phys.Mass.Techn.Inst.$\underline {2},\,(3)\,,131\,,
\,(1923)\,;$\\
Proc.Mathm.Soc.(London),$\underline{22} ,\,(6),\,457,\,(1924)\,.$\\
17.Feynman R.P.,Review Mod.Phys.$\,\underline{20} ,\,(2),\,367,\,(1948)\,;
$ \\
Phys.Review ,$\underline{76} ,\,(6),\,769\,,\,(1948)\,,\underline{84} ,\,
(1),\,108,\,(1951)\,.$\\
18.Schrodinger E.,Sitzunsber.Preuss.Akad.Wiss.,$ K1, 418 (1930) $\\
Berlin.Bericht.$\,296,\,400,\,(1930)\,;\,144,\,(1931)\,.$\\
19.Pauli W.,Handbuch der Physik,$\underline{24} ,\,(1),\,210,\,(1933)\,,$
Springer,Berlin \\
20.Welton Th.,Phys.Review $\underline{74} ,1157,(1948).$\\
21.Fenyes I.,Ztschr.f.Phys.$\underline{132} ,81,(1952).$\\
22.Bohm D., Phys. Review ,$\underline{85} ,166, 180, (1952).$\\
23.Rangelov J.M.,\,Report Series of Symposium on the Foundations \\
of Modern Physics,\,6/8,August,1987,\,Joensuu,95-99,FTL,131,Turqu ,Finland\\
24.Rangelov J.M.,Problems in Quantum Physics 2,Gdansk 89,18-23 \\
September 1989 Gdansk, p.461-483,World Scientific,Singapure,1990 .\\
25.Rangelov J.M., Abstracts Booklet of 29th Annual Conference of the \\
University of Peoples' Friendship ,Moscow 17-31 may 1993,Physical ser.\\
26.Rangelov J.M., Abstracts Booklet of Symposium on the Foundations \\
of Modern Physics, 13/16 ,June , 1994 ,Helsinki , Finland 60-62 .\\
27.Sokolov A.A.,Scientific reports of higher school,(1),120.(1950).
Moscow ;\\
Philosophical problems of elementary particle physics.\, Acad.of Scien.\\
of UdSSR ,Moscow ,188, (1963) .\\
28.Rangelov J.M.,Abstract Booklet of B R U - 2\,,\,12-14\,September,
1994, \\
Ismir\,,Turkey ;$\quad$ Balk.Phys.Soc.$\underline{2},(2),1974,(1994).$\\
29.Rangelov J.M.,Abstract Booklet of B R U - 3 \,,\,2-5 September, 1997, \\
Cluj-Napoca,Romania .\\

\end{document}